\documentclass[11pt]{article}
\usepackage{latexsym}
\usepackage{amsmath} 
\usepackage{epsfig}
\usepackage{floatfig}
\begin{document}
\initfloatingfigs

\title{{\bf Screw Photon-like (3+1)-Solitons in Extended Electrodynamics\\
            }}
\author{{\bf Stoil Donev}\\ Institute for Nuclear Research and Nuclear
Energy,\\ Bulg.Acad.Sci., 1784 Sofia, blvd.Tzarigradsko chausee 72\\
Bulgaria\\ e-mail: sdonev@inrne.bas.bg,\\}

\date{}

\maketitle

\begin{abstract}
This paper aims to present explicit photon-like (3+1) spatially finite
soliton solutions of screw type to the vacuum field equations of Extended
Electrodynamics (EED) in relativistic formulation. We begin with emphasizing
the need for spatially finite soliton modelling of microobjects. Then we
briefly comment the properties of solitons and photons and recall some facts
from EED.  Making use of the localizing functions from differential topology
(used in the partition of unity) we explicitly construct spatially
finite screw solutions.  Further a new description of the spin momentum
inside EED, based on the notion for energy-momentum exchange between $F$ and
$*F$, is introduced and used to compute the integral spin momentum of a screw
soliton.  The consistency between the spatial and time periodicity naturally
leads to a particular relation between the longitudinal and transverse sizes
of the screw solution, namely, it is equal to $\pi$.  Planck's formula
$E=h\nu$ in the form of $ET=h$ arizes as a measure of the integral spin
momentum.

\end{abstract}

\section{Introduction}

The very notion of {\it really existing objects}, i.e. {\it physical objects
carrying energy-momentum}, necessarily implies that all such objects must
have definite stability properties, as well as propertiies that do not change
with time; otherwise everything would constantly change and we could not talk
about objects and structure at all, moreover no memory and no knowledge would
be possible.  Through definite procedures of {\it measurement} we determine,
where and when this is possible, quantitative characteristics of the
physical objects.  The characteristics obtained differ in: their nature and
qualities; in their significance to understand the structure of the objects
they characterize; in their abilities to characterize interaction among
objects; and in their universality.

Natural objects may be classified according to various principles.  The
classical point-like objects (called usually particles) are allowed to
interact continuously with each other just through exchanging (through some
mediator usually called field) universal conserved quantities: energy,
momentum, angular momentum, so that, the set of objects "before" interaction
is the same as the set of objects "after" interaction, no objects have
disapeared and no new objects have appeared, only the conserved quantities
have been redistributed. This is in accordance with the assumption of
point-likeness, i.e.  particles are assumed to have no internal structure, so
they are undestroyable. Hence, classical particles may be subject only to
elastic interaction.

Turning to study the set of microobjects, called usually elementary
particles:  photons, electrons, etc., physicists have found out, in contrast
to the case of classical particles, that a given set of microobjects may
transform into {\it another} set of microobjects under definite conditions,
for example, the well known anihilation process:  $(e^{+},e^{-})\rightarrow
2\gamma$. These transformations obey also the energy-momentum and
angular+spin momentum conservation, but some features may disapear (e.g. the
electric charge) and new features (e.g. motion with the highest velosity) may
appear.  Hence, microobjects {\it allow to be destroyed}, so they have
structure and, consequently, they would NOT admit the approximation
"point-like objects". In view of this we may conclude that any theory aiming
to describe their behaviour must take in view their structure. In particular,
assuming that the Planck formula $E=h\nu$ is valid for a free microobject,
the only reasonable way to understand where the characteristic {\it
frequency} comes from is to assume that this microobject has a periodic
dynamical structure.

The idea of conservation as a dynamics generating rule was realized and
implemented in physics firstly by Newton a few centuries ago in the frame of
real (classical) bodies: he invented the quantity {\it momentum}, ${\bf p}$,
for an isolated body (treated as a point, or structureless object), as its
integral characteristic, postulated its time-constancy (i.e. conservation),
and wrote down his famous equation $\dot {\bf p}$=${\bf F}$.  This equation
says, that the integral characteristic {\it momentum} of a given point-like
object may (smoothly) increase {\it only} if some other object loses
(smoothly) the same quantity of {\it momentum}.  The concept of {\it force},
${\bf F}$, measures the momentum transfered per unit time.  The
conservative quantities {\it energy} and {\it angular momentum}, consistent
with the momentum conservation, were also appropriately incorporated.  The
most {\it universal} among these turned out to be {\it the energy}, since, as
far as we know, every natural object carries energy. This property of
universality of energy makes it quite distinguished among the other
conservative quantities, because its change may serve as a relyable measure
of {\it any} kind of interaction and transformation of real objects.

We especially note, that all these conservative quantities are carried by
some object(s), {\it no energy-momentum can exist without corresponding
objects-carryers}. In this sense, the usual words "energy quanta" are
sensless if the corresponding carryers are not pointed out.

So, theoretical physics started with idealizing the natural objects as
particles, i.e. objects {\it without structure}, and the real world was
theoreticaly viewed as a collection of {\it interacting}, i.e.
energy-momentum exchanging, particles. As far as the behaviour of the real
objects as a whole is concerned, and the interactions considered do NOT lead
to destruction of the bodies-particles, this theoretical model of the real
world worked well.

The 19th century physics, due mainly to Faraday and Maxwell, created the
theoretical concept of {\it electromagnetic field} as the {\it interaction
carrying object}, responsible for the observed mutual influence between
distant electrically charged (point-like) objects. This concept presents the
electromagnetic field as an extended (in fact, infinite) continuous object,
having dynamical structure, and although it influences the behaviour of the
charged particles, it does NOT destroy them.  The theory of the
electromagnetic field was based also on balance relations  of new kind of
quantities. Actually, the new concepts of {\it flux of a vector field through
a 2-dimensional surface} and {\it circulation of a vector field along a
closed curve} were coined and used extensively. The Faraday-Maxwell equations
in their integral form establish, in fact, where the time-changes of the
fluxes of the electric and magnetic fields go to, or come from, in both cases
of a {\it closed} 2-surface, and of a {\it not-closed} 2-surface with a
boundary, and in this sense they introduce a kind of {\it ballance
relations}. We note, that these fluxes are new quantities, specific to the
 continuous character of the physical object under consideration; the field
equations of Faraday-Maxwell do NOT express directly energy-momentum balance
relations as the above mentioned Newton's law ${\dot {\bf p}}={\bf F}$ does.
Nevertheless, they are consistent with energy-momentum conservation, as it is
well known.  The corresponding local energy-momentum quantities turn out to
be quadratic functions of the electric and magnetic vectors.

Although very useful for considerations in finite regions with boundary
conditions, the pure field Maxwell equations have time-dependent solutions in
the whole space that could hardly be considered as mathematical models of
really existing fields. As a rule, if these solutions are time-stable and not
static, they occupy the whole 3-space, or its infinite sub-region (e.g. plane
waves), and, hence, they carry {\it infinite} energy and momentum (infinite
objects).  On the other hand, according to Cauchy's theorem for the
D'Alembert wave equation [1], which is necessarily satisfied by any component
of the vacuum field in Maxwell theory, every {\it finite} (and smooth enough)
initial field configuration is strongly time-unstable: the initial
condition blows up radially and goes to infinity, and its forefront and
backfront propagate with the velocity of light.  Hence, Faraday-Maxwell
equations {\it cannot describe finite time-stable} localized
field-objects.

The inconsistences between theory and experiment appeared in a full scale at
the end of the last century and it soon became clear that they were not
avoidable in the frame of classical physics.  After Planck and Einstein
created the notion of {\it elementary field quanta}, named later by Lewis [2]
{\it photon}, physicists faced the above mentioned problem:  the light quanta
appeared to be real objects of a new kind, namely, they did NOT admit the
point-like approximation like Newton's particles did. In fact, every photon
propagates (transitionally) as a whole with a {\it constant} velocity and
keeps unchanged the energy-momentum it carries, which {\it should} mean that
it is a {\it free} object.  On the other hand, it satisfies Planck's relation
$E=h\nu$, which means that the very existence of photons is intrinsically
connected with a periodical process of frequency $\nu$, and periodical
processes in classical physics are generated by {\it external} force-fields,
which means that the object {\it should not} be free.  The efforts to
overcome this undesirable situation resulted in the appearence of quantum
theory , quantum electrodynamics was built, but the assumption "the
point-like approximation works" {\it was kept as a building stone} of the new
theory, and this braught the theory to some of the well known singularity
problems.

Modern theory tries to pay more respect to the view that the point-like
approximation does NOT work in principle in the set of microobjects
satisfying the above Planck's formula. In other words, the right
theoretical notion for these objects should be that of {\it extended
continuous finite} objects, or {\it finite field objects}.
During the last 30 years physicists have been trying seriously
to implement in theory the "extended point of view" on microobjects mainly
through the string/brane theories, but the difficulties these theories
meet, generated partly by the great purposes they set before themselves, still
do not allow to get satisfactory results. Of course, attempts to create
extended point view on elementary particles different from the string/brane
theory approach, have been made and are beeng made these days [3].

Anyway, we have to admit now, that after one century away from the discovery
of Planck's formula, we still do not have a complete and satisfactory
self-consistent theory of single photons. So, creation of a self-consistent
extended point of view and working out a corresponding theory is still a
challenge,  and this paper aims to consider such an extended point of view on
photons as {\it screw solitons} in the frame of the newly developed EED [4].
First we summurize the main features/properties of solitons and photons.

\section{Solitons and Photons}
The concept of soliton appears in physics as a nonlinear elaboration -
physical and mathematical- of the general notion for exitation in a medium.
It includes the following features:
\vskip 0.4cm
I. PHYSICAL.
\vskip 0.4cm 1.
The medium is homogeneous, isotropic and has definite properties of
elasticity.
\vskip 0.3cm
2. The exitation does not destroy the medium.
\vskip 0.3cm
	3. The exitation is finite:
\vskip 0.3cm
-at every moment it occupies comparetively small volume of the medium,

-it carries finite quantities of energy-momentum and angular
momentum, and of any other physical quantity too,

-it may have translational-rotational (may time-periodical) dynamical
structure, i.e.  besides its straightline propagation as a whole it may have
internal rotational degrees of freedom.
\vskip 0.3cm
	4. The exitation is time-stable, i.e. at lack of external
perturbations its dynamical evolution does not lead to a self-ruin.
In particular, the spatial shape of the exitation does not (significantly)
change during its propagation.
\vskip 0.3cm
The above 4 features outline the physical notion of a {\it solitary wave}. A
solitary wave becomes a {\it soliton} if it has in adition the following
property of stability:
\vskip 0.3cm
	5. The exitation survives when collides with another exitation of the
same nature.
\vskip 0.3cm
We make some comments on the features 1-5.

Feature 1 requires homogenity and some ealstic properties of the medium,
which means that it is capable to bear the exitation, and every region of it,
subject to the exitation, i.e. draged out of its natural (equilibrium) state,
is capable to recover entirely after the exitation leaves that region.

Feature 2 puts limitations on the exitations considered in view of the medium
properties:  they should not destroy the medium.

Feature 3 is very important, since it requires {\it finite} nature of the
exitations, it enables them  to represent some initial level self-organized
physical objects with dynamical structure, so that these objects "feel good"
in this medium. This finite nature assumption admits only such exitations
which may be {\it created} and {\it destroyed}; no point like and/or
infinite exitations are admitted.  The exitation interacts permanently with
the medium and the time periodicity available may be interpreted as a measure
of this interaction.

Feature 4 guarantees the very existence of the exitation in this medium, and
the shape keeping during propagation allows its recognition and
identification when observed from outside. This feature 4 carries in some
sense the first Newton's principle from the mechanics of particles to the
dynamics of continuous finite objects, it implies conservation of
energy-momentum and of the other characteristic quantities of the exitation.

The last feature 5 is frequently not taken in view, especially when one
considers single exitations. But in presence of many exitations in a given
region it allows only such kind of interaction between/among the exitations,
which does not destroy them, so that the exitations get out of the
interaction (almost) the same. This feature is some continuous version of the
elastic collisions of particles.
\vskip 0.4cm
II. MATHEMATICAL
\vskip 0.4cm
1. The exitation defining functions $\Phi^a$ are components of {\it one}
mathematical object (usually a section of a vector/tensor bundle) and depend
on $n$ spatial and $1$ time coordinates.
\vskip 0.3cm
2. The components $\Phi^a$ satisfy some system of nonlinear partial
differential equations (except the case of (1+1) linear wave equation), and
admit some "running wave" dynamics as a whole, together with available
internal dynamics.
\vskip 0.3cm
3. There are (infinite) many conservation laws.
\vskip 0.3cm
4. The components $\Phi^a$ are localized (or finite) functions with respect
to the spatial coordinates, and the conservative quantities are finite.
\vskip 0.3cm
5. The multisoliton solutions, describing elastic interaction (collision),
tend to many single soliton solutions at $t\rightarrow \infty$.
\vskip 0.4cm
Comments:
\vskip 0.3cm
1. Feature 1 introduces some notion of {\it integrity}: {\it one exitation -
one mathematical object}, although having many algebraically independant but
differentially interrelated (through the equations) components $\Phi^a$.
\vskip 0.3cm
2. Usually, the system of PDE is of {\it evolution} kind, so that the
exitation is modelled as a dynamical system: the initial configuration
determines fully the evolution. The "running wave" dynamics as a whole
introduces Galileo/Lorentz invariance and corresponds to the physical
feature 4. The nonlinearity of the equations is meant to guarantee the
spatially localized (finite) nature of the solutions.
\vskip 0.3cm
3. The infinite many consrvation laws frequently lead to complete
integrability of the equations.
\vskip 0.3cm
4. The spatially localized nature of $\Phi^a$ represents the finite nature of
the exitation.
\vskip 0.3cm
5. The many single soliton asymptotics at $t\rightarrow\infty$ of a
multisoliton solution mathematically represents the elastic character of the
interactions admitted, and so it takes care of the stability of the
physical objects beeing modelled.

\vskip 0.4cm
The above physical/mathematical features are not always strictly accounted
for in the literature. For example, the word {\it soliton} is frequently used
for a {\it solitary wave} exitation.  Another example is the usage of the
word soliton just when the energy density, being usually a quadratic function
of the corresponding $\Phi^a$, has the above soliton properties [5]. Also,
one usually meets this soliton terminology for spatially {\it localized},
i.e.  going to zero at spatial infinity, but {\it not} spatially {\it finite}
$\Phi^a$, i.e. when the spatial support of $\Phi^a$ is a compact set. In
fact, all soliton solutions of the well known KdV, SG, NLS equations are
localized and {\it not} finite.  It is curious, that the {\it linear} (1+1)
wave equation has spatially finite soliton solutions of arbitarary shape.

Further in this paper we shall present 1-soliton screw solutions of the
vacuum EED equations, so we may, and shall, use the more attractive word
soliton for solitary wave. We hope this will not bring any troubles to the
reader.

The screw soliton solutions we are going to present are of photon-like
character, i.e. the velocity of their translational component of propagation
is equal to the velocity of light $c$, and besides of the energy-momentum,
they carry also internal angular (spin, helicity) momentum accounting for the
available rotational component of propagation. Therefore, this seems to be
the proper place to recall some of the well known properties of {\it
photons}.

First of all, as it was explained in the Introduction, photons should not be
considered as point-like objects since they respect the Planck's relation
$E=h\nu$ and carry internal (spin) momentum, so they have to be considered as
extended finite objects with periodical rotational-translational dynamical
sructure.  Therefore, we assume that the concept of soliton, as described
above, may serve as a good mathematical tool in trying to represent
mathematically the real photons, so that, their well known integral
properties to appear as determined by their dynamical structure.

We give now some of the more important for our purposes properties of
photons.
\vskip 0.3cm
	1. Photons have zero proper mass and electric charge. The
(straightline) translational component of their propagation velocity is
constant and equal to the experimentally established velocity of light in
vacuum.
\vskip 0.3cm
	2. Photons are time-stable objects. Every interaction with other
objects kills them.
\vskip 0.3cm
	3. The existence of photons is generically connected with some
time-periodical process of period $T$ and frequency $\nu=T^{-1}$, so that
the Planck relation $E=h\nu$, or $ET=h$, where $h$ is the Planck constant,
holds.
\vskip 0.3cm
	4. Every single photon carries momentum ${\bf p}$ with
$|{\bf p}|=h\nu/c$ and spin momentum equal to the Planck constant $h$.
\vskip 0.3cm
	5. Photons are polarized objects. The polarization of every single
photon is determined through the relation between the translational and
rotational directions of propagation, hence, the 3-dimensionality of the real
space allows just two polarizations. We call the polarization {\it right}, if
when looking along its translational component of propagation, i.e. from
befind, we find its rotational component of propagation to be {\it
clock-wise}, and the polarization is {\it left} when under the same
conditions we find {\it anti-clock-wise} rotational component of propagation.
\vskip 0.3cm
6. Photons do not interact with each other. i.e. they pass through each other
without changes.
\vskip 0.4cm
These well known properties of photons would hardly need any comments.
However, according to our oppinion, these properties strongly suggest to
make use of the soliton concept for working out a mathematical model of their
structure and propagation. And this was one of the main reasons to develop
the extension of Faraday-Maxwell theory to what we call now Extended
Electrodynamics (EED). We proceed now to recall the basics of EED, in the
frame of which the screw soliton model of photons will be worked out.

\section{Basics of Extended Electrodynamics in Vacuum}

We are going to consider just the vacuum case of EED in relativistic
formulation.  The signiture of the space time pseudometric $\eta$ is
$(-,-,-,+)$, the canonical coordinates will be denoted by
$(x^1,x^2,x^3,x^4)=(x,y,z,\xi=ct)$, so the components $\eta_{\mu\nu}$ in any
canonical coordinate system are
$-\eta_{11}=-\eta_{22}=-\eta_{33}=\eta_{44}=1$, and $\eta_{\mu\nu}=0$ for
$\mu\neq\nu$. The corresponding volume 4-form $\omega_o$ is given by
$\omega_o=dx\wedge dy\wedge dz\wedge d\xi$ since $|det(\eta_{\mu\nu})|=1$.
The Hodge $*$-operator is defined by
$\alpha\wedge \beta=\eta(*\alpha,\beta)\omega_o$, where $\alpha$ and $\beta$
are $p$ and $4-p$ forms respectively.
In terms of $\varepsilon_{\mu\nu\sigma\rho}$ we have
$(*F)_{\mu\nu}=-\frac12\varepsilon_{\mu\nu\sigma\rho}F^{\sigma\rho}$. We have
also the exterior derivative ${\bf d}$ and the coderivative
$\delta=*{\bf d}*$. The physical interpretation of $F_{\mu\nu}$ are:
$F_{i4}=-F_{4i}, i=1,2,3$, are the components ${\bf E}^1, \mathbf{E}^2,
\mathbf{E}^3$ of the electric vector ${\bf E}$, and $(F_{23},-F_{13},F_{12})$
are the components ${\bf B}^1, {\bf B}^2, {\bf B}^3$ of the
magnetic vector ${\bf B}$, respectively.

In terms of $\delta$ the vacuum Maxwell equations are given by
\begin{equation}
\delta*F=0,\ \ \ \delta F=0.                            
\end{equation}
In EED the above equations (1) are extended to
\begin{equation}
\delta*F\wedge F=0,\ \ \delta F\wedge *F=0,\ \          
\delta F\wedge F- \delta*F\wedge*F=0.
\end{equation}
In components, equations (2) are respectively:
\begin{equation}
(*F)_{\mu\nu}(\delta*F)^\nu=0,\ \ F_{\mu\nu}(\delta F)^\nu=0,\ \
(*F)_{\mu\nu}(\delta F)^\nu+F_{\mu\nu}(\delta*F)^\nu=0.            
\end{equation}
The Maxwell energy-momentum tensor
\begin{equation}
Q_\mu^\nu=-\frac{1}{8\pi}\big[F_{\mu\sigma}F^{\nu\sigma}+        
(*F)_{\mu\sigma}(*F)^{\nu\sigma}\big]
\end{equation}
is assumed as energy-momentum tensor in EED because its divergence
\begin{equation}
\nabla_{\nu}Q^\nu_\mu=\frac{1}{4\pi}\big[F_{\mu\nu}(\delta F)^\nu+   
(*F)_{\mu\nu}(\delta*F)^\nu\big]
\end{equation}
is obviously zero on the solutions of equations (3) and no problems of
Maxwell theory at this point are known. The physical sense of
equations (3) is, obviously, local energy-momentum redistribution during the
time-evolution: the first two equations say that $F$ and $*F$ keep locally
their energy-mimentum, and the third equation says (in correspondence with
the first two), that the energy-momentum transferred from $F$ to $*F$ is
always equal locally to the energy-momentum transferred from $*F$ to $F$,
hence, any of the two expressions $F_{\mu\nu}(\delta *F)^\nu$ and
$(*F)_{\mu\nu}(\delta F)^\nu$ may be considered as a measure of the
rotational component of the energy-momentum redistribution between $F$ and
$*F$ during propagation (recall that the spatial part of $\delta F$ is ${\rm
rot}{\bf B}$ and the spatial part of $\delta*F$ is ${\rm rot}{\bf E}$).

Obviously, all solutions of (1) are solutions to (3), but equations (3) have
more solutions. In particular, those solutions of (3) which satisfy the
relations
\begin{equation}
\delta F\neq 0, \ \delta*F\neq 0                          
\end{equation}
are called {\it nonlinear}.
Further we are going to consider only the nonlinear solutions of (3).

Some of the basic results in our previous studies of the nonlinear solutions
of equations (3) could be summarized in the following way:
For every nonlinear solution $(F,*F)$ of
(3) there exists a canonical system of coordinates $(x,y,z,\xi)$ in which the
solution is fully represented by two functions $\Phi(x,y,\xi+\varepsilon z)$,
$\varepsilon=\pm 1$,
and $\varphi(x,y,z,\xi),\ |\varphi|\leq 1$, as follows:
\[
F=\varepsilon \Phi\varphi dx\wedge dz + \Phi\varphi dx\wedge d\xi +
\varepsilon \Phi\sqrt{1-\varphi^2}dy\wedge dz +
\Phi\sqrt{1-\varphi^2}dy\wedge d\xi,
\]
\[
*F=-\Phi\sqrt{1-\varphi^2}dx\wedge dz -
\varepsilon \Phi\sqrt{1-\varphi^2}dx\wedge d\xi +
\Phi\varphi dy\wedge dz + \varepsilon \Phi\varphi dy\wedge d\xi.
\]
We call $\Phi$ the {\it amplitude} function and $\varphi$ the {\it phase}
function of the solution. The condition $|\varphi|\leq 1$ allows to set
$\varphi=
{\rm cos}\psi$, and further we are going to work with $\psi$, and
$\psi$ will be called {\it phase}. As we showed [4], the two functions $\Phi$
and $\varphi$ may be introduced in a coordinate free manner, so they have
well defined invariant sense. Every nonlinear solution satisfies the
following important relations:
$$
(\delta F)^2<0,\ (\delta*F)^2<0,\
|\delta F|=|\delta *F|, \ (\delta F)_\sigma(\delta *F)^\sigma=0, \
\ F_{\mu\nu}F^{\mu\nu}=F_{\mu\nu}(*F)^{\mu\nu}=0.
$$
We recall also the {\it scale factor} $L$, defined by the relation
$L=|\Phi|/|\delta F|$. A simple calculation shows that it depends only on the
derivatives of $\psi$ in these coordinates and is given by
\begin{equation}
L=\frac{1}{|\psi_\xi-\varepsilon\psi_z|}.                   
\end{equation}

\section{Screw Soliton Solutions in Extended Electrodynamics}

Note that EED considers the field as having two components: $F$ and $*F$.
As we mentioned earlier, the third equation of (3) describes how much
energy-momentum is redistributed locally with time between the two components
$F$ and $*F$ of the field: $F_{\mu\nu}(\delta *F)^\nu dx^\mu$ gives the
transfer from $F$ to $*F$, and $(*F)_{\mu\nu}\delta F^\nu dx^\mu$ gives the
transfer from $*F$ to $F$, thus, if there is such an energy-momentum exchange
equations (3) require permanent and equal mutual energy-momentum transfers
between $F$ and $*F$.  Since $F$ and $*F$ are always orthogonal to each other
\newline
[$F_{\mu\nu}(*F)^{\mu\nu}=0$] and these two mutual transfers depend on the
derivatives of the field functions through $\delta F$ and $\delta *F$ (i.e.
through ${\rm rot}\mathbf{B}$ and ${\rm rot}\mathbf{E}$, which are not equal
to zero in general), we may interpret this property of the solution as a
description of an {\it internal rotation-like} component of the general
dynamics of the field.  Hence, any of the two expressions $F_{\mu\nu}(\delta
*F)^\nu dx^\mu$ or $(*F)_{\mu\nu}\delta F^\nu dx^\mu$, having the sense of
local energy-momentum change, may serve as a natural measure of this
rotational component of the energy-momentum redistribution during the
propagation.  Therefore, after some appropriate normalization, we may
interpret any of the two 3-forms $(*F)\wedge (\delta *F)$ and
$F\wedge \delta F$ as local {\it spin-momentum} of the solution.
Making use of the above expressions for $F$ and $*F$ we compute $F\wedge
\delta F=(*F)\wedge (\delta *F)$:
\[
F\wedge \delta F=-\varepsilon
\Phi^2(\psi_\xi-\varepsilon \psi_z)dx\wedge dy\wedge
dz-\Phi^2(\psi_\xi-\varepsilon\psi_z)dx\wedge dy\wedge d\xi.
\]
Since $\Phi\neq 0$, this expression says that we shall have nonzero local
spin-momentum only if $\psi$ is {\it not} a running wave along $z$.  The
above idea to consider the 3-form $F\wedge \delta F$ as a measure of the spin
momentum of the solution suggests also some additional equation for $\psi$,
because the spin momentum is a conserved quantity and its integral over the
3-space should not depend on the time variable $\xi$.  This requires to have
some nontrivial {\it closed 3-form} on ${\cal R}^4$, such that when
restricted to the 3-space and (spatially) integrated to give the integral
spin momentum of the solution. Therefore we assume the additional equation
\begin{equation}
\mathbf{d}(F\wedge \delta F)=0.  
\end{equation}
In our system of coordinates this equation is reduced to
\begin{equation}
\mathbf{d}(F\wedge \delta F)=\varepsilon\Phi^2\left(\psi_{\xi\xi}+\psi_{zz}-
2\varepsilon\psi_{z\xi}\right)dx\wedge dy\wedge dz\wedge d\xi=0,   
\end{equation}
i.e.
\begin{equation}
\psi_{\xi\xi}+\psi_{zz}-2\varepsilon\psi_{z\xi}=
\left(\psi_\xi-\varepsilon\psi_z\right)_\xi -
\varepsilon\left(\psi_\xi-\varepsilon\psi_z\right)_z =0.           
\end{equation}
Equation (10) has the following solutions:

1$^o $. Running wave solutions $\psi=\psi(x,y,\xi+\varepsilon z)$,

2$^o $. $\psi = \xi.g(x,y,\xi+\varepsilon z)+b(x,y)$,

3$^o $. $\psi = z.g(x,y,\xi+\varepsilon z)+b(x.y)$,

4$^o $. Any linear combination of the above solutions with coefficients which
are allowed to depend on $(x,y)$.
The functions $g(x,y,\xi+\varepsilon z)$ and $b(x,y)$ are arbitrary in the
above expressions.

The running wave solutions $\psi_1$, defined by $1^o$ lead to $F\wedge \delta
F=0$ and to $|\delta F|=0$, and by this reason they have to be ignored. The
solutions $\psi_2$ and $\psi_3$, defined respectively by 2$^o$ and 3$^o$,
give the same scale factor $L=1/|g|$.  Since at all spatial points
where the field is different from zero we have $\xi+\varepsilon
z=const$, we may choose $|g(x,y,\xi+\varepsilon z)|=1/l(x,y)>0$, so we obtain
the following {\it nonrunning} wave solutions of (10):
\begin{equation}
\psi_2=\frac{\kappa \xi}{l(x,y)}+b(x,y);
\ \ \psi_3=\frac{\kappa z}{l(x,y)}+b(x,y),                  
\end{equation}
where $\kappa=\pm 1$ accounts for the two different polarizations.  Clearly,
the physical dimension of $l(x,y)$ is {\it length}, $b(x,y)$ is dimensionless
and the {\it scale} factor is $L=l(x,y)$.

The corresponding electric $\mathbf{E}$ and magnetic $\mathbf{B}$
vectors for the case 2$^o$ in view of (11) are
\begin{equation}
\mathbf{E}=\Big[\Phi(x,y,\xi+\varepsilon z){\rm
cos}\left(\frac{\pm \xi}{l(x,y)}+b(x,y)\right); \Phi(x,y,\xi+\varepsilon
z){\rm sin}\left(\frac{\pm \xi}{l(x,y)}+b(x,y)\right); 0\Big],      
\end{equation}
\begin{equation}
\mathbf{B}=\Big[\varepsilon\Phi(x,y,\xi+\varepsilon z){\rm
sin}\left(\frac{\pm \xi}{l(x,y)}+b(x,y)\right);-\varepsilon
\Phi(x,y,\xi+\varepsilon z){\rm cos}\left(\frac{\pm
\xi}{l(x,y)}+b(x,y)\right); 0\Big],                                    
\end{equation}
A characteristic feature of the solutions, defined by (12)-(13), is that the
direction of the electric and magnetic vectors at some initial moment $t_o$,
as seen from (12)-(13), is entirely determined by $(\psi_2)_{t_o}$ (under
$L=1/|g|=l(x,y))$, and so, it does not depend on the coordinate $z$.
Therefore, at all points of the 3-region $\Omega_o$, occupied by the solution
at $t_o$, under the additional conditions $L=l(x,y)=const$ and
$b(x,y)=const$, the directions of the representatives of $\mathbf{E}$ at all
spatial points will be the same, independently on the spatial shape of the
3-region $\Omega_o$. At every subsequent moment this common direction will be
rotationally displaced, but will stay the same for the representatives of
$\mathbf{E}$ at the different spatial points.  The same feature will hold for
the representatives of $\mathbf{B}$ too. Thus, the representatives of the
couple $(\mathbf{E},\mathbf{B})$ will rotate in time, clockwise ($\kappa=-1$)
or anticlockwise ($\kappa=1$), coherently at all spatial points, so these
solutions show no twist-like, or screw, propagation component even if the
region $\Omega_o$ has a screw shape.  Hence, these solutions may be
soliton-like but not of screw kind.

The electric $\mathbf{E}$ and magnetic $\mathbf{B}$ vectors for the case
3$^o$ in view of (11) are
\begin{equation}
\mathbf{E}=\Big[\Phi(x,y,\xi+\varepsilon z){\rm
cos}\left(\frac{\pm z}{l(x,y)}+b(x,y)\right); \Phi(x,y,\xi+\varepsilon z){\rm
sin}\left(\frac{\pm z}{l(x,y)}+b(x,y)\right); 0\Big],      
\end{equation}
\begin{equation}
\mathbf{B}=\Big[\varepsilon\Phi(x,y,\xi+\varepsilon z){\rm
sin}\left(\frac{\pm z}{l(x,y)}+b(x,y)\right);-\varepsilon
\Phi(x,y,\xi+\varepsilon z){\rm cos}\left(\frac{\pm z}{l(x,y)}+b(x,y)\right);
0\Big],                                                    
\end{equation}
We are going to use this particular solution  (14)-(15) to construct a
theoretical example of a screw photon-like soliton solution. We have to give
an explicit form of the amplitude function $\Phi$. In accordance with the
{\it finite} nature of the solution $\Phi$ must have compact spatial support,
and this guarantees that the solution is finite because the phase function
$\varphi={\rm cos}$, so, the products $\Phi.{\rm cos}(\psi)$, $\Phi.{\rm
sin}(\psi) $ are also finite.
\par
\begin{floatingfigure}{25em}
\mbox{\epsfig{file=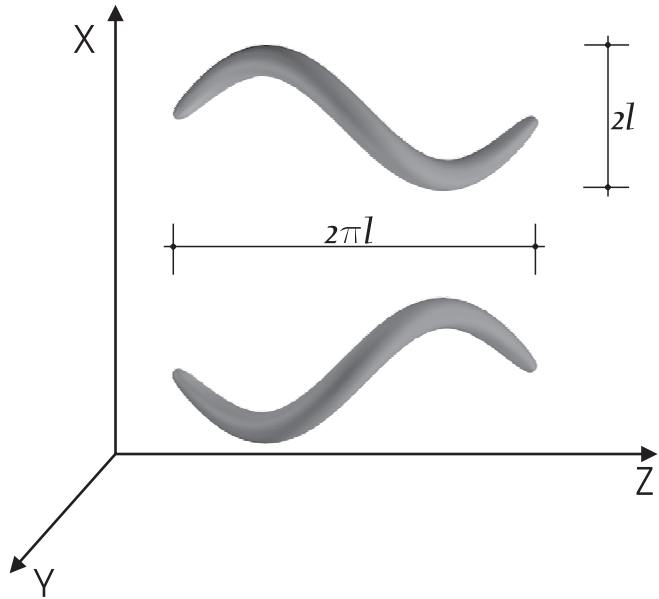}}
 \caption{ An animation, visualizing the 2-kind polarized screw
 photon-like soliton solutions
 \label{fig:FIG1}}
\end{floatingfigure}  \quad

First we outline the idea (FIG1).  We mean to choose $\Phi=\Phi_o$ at $\xi=0$
in such a way that $\Phi_o(x,y,z)$ to be localized inside a screw cylinder of
radius $r_o$ and height $|z_2-z_1|=2\pi l_o$ and we shall denote this region
by $\Omega_o$; the $z$-prolongation of $\Omega_o$ will be an infinite screw
cylinder denoted by $\Omega$.  Let this screw cylinder have the coordinate
axis $z$ as an outside axis, i.e. $\Omega$  {\it windes around} the $z$-axis
and {\it never} crosses it.  The initial configuration $\Omega_o$ shall be
made propagating along $\Omega$ through both: $z$-tanslation, and a
consistent rotation around itself and around the $z$-axis.  This kind of
propagation will be achieved only if every spatial point inside $\Omega_o$
will keep moving along its own screw line and will never cross its
neighbors' screw lines.

We consider the plane $(x,y)$ at $z=0$ and choose a point $P=(a,a), a>0$ in
the first quandrant in this plane, so that the points $(x,y,z=0)$ where
$\Phi_o(x,y,z=0)\neq 0$ are inside the circle $x^2+y^2\leq r_o^2$
centered at $P=(a,a)$, and the distance $a\sqrt{2}$ between $P$ and the
zero-point $(0,0,0)$ is much greater than $r_o$.

Now we consider the function
\[
\frac{1}{{\rm cosh}\left[
\left(x-a)\right)^2+
\left(y-a)\right)^2
\right]}.
\]
It is concentrated mainly inside $\Omega$ and goes to zero outside $\Omega$
although it becomes zero just at infinity.  Restricted to the plane $z=0$ it
is concentrated inside the circle $A_{r_o}$ of radius $r_o$ around the point
$P$ and it becomes zero at infinity ($x=\infty, y=\infty$).  We want this
concentrated but {\it still infinite} object to become {\it finite}, i.e. to
become zero outside $\Omega$.  In order to do this we shall make use of the
so called {\it localizing} functions, which are smooth everywhere, are equal
to 1 on some compact set $A$ and quickly go to zero outside $A$.  (These
functions are very important in differential topology for making partition of
unity and glueing up various structures [6]). We shall denote these functions
by $\theta(x,y,....)$.  Let $\theta(x,y;r_o)$ be a localizing function around
the point $P$ in the plane $(x,y)$, such that it is equal to $1$ inside the
circle $A_{r}$, centered at $P$, of radius $r$, and $r$ is considered to be a
bit shorter but nearly equal to $r_o$, and $\theta(x,y;r_o)=0$ outside the
circle $A_{r_o}$.  We modify now the above considered function as follows:
\[
\frac{\theta(x,y;r_o)}
{{\rm cosh}\left[\left(x-a\right)^2+
\left(y-a\right)^2\right]}.
\]
Let now $\theta(z;l_o)$ be a localizing function with respect to the interval
$(z_1,z_2), z_2>z_1$, where $|z_2-z_1|=2\pi l_o$, Hence, the function
\[
\Phi_o(x,y,0+z)=\frac{\theta(x,y;r_o)\theta(z;l_o)}
{{\rm cosh}\left[\left(x-a\right)^2+
\left(y-a\right)^2\right]}
\]
is different from zero only inside $\Omega_o$. We choose now the scale factor
$l(x,y)$ to be a constant equal to $l_o$, and making use of the phase
$\psi_3$ with $\kappa=1$ we define the function
\begin{equation}
\Phi_o\psi_3^+=C\frac{\theta(x,y;r_o)\theta(z;l_o)}
{{\rm cosh}\left[\left(x-a\right)^2+
\left(y-a\right)^2\right]}                    
{\rm cos}\left(\frac{z}{l_o}+b(x,y)\right),
\end{equation}
where $C$ is a constant with appropriate physical dimension. This function
(16) represents the first component ${\bf E}_1$ of the electric vector at
$t=0$.  Similarly, for ${\bf E}_2$ at $t=0$ we write
\begin{equation}
\Phi_o\psi_3^+=C\frac{\theta(x,y;r_o)\theta(z;l_o)}
{{\rm cosh}\left[\left(x-a\right)^2+
\left(y-a\right)^2\right]}                    
{\rm sin}\left(\frac{z}{l_o}+b(x,y)\right),
\end{equation}
The right hand sides of (16) and (17) define the initial state of the
solution, and this initial condition occupies the screw cylinder $\Omega_o$,
its internal axis is a screw line away from the $z$-axis at a distance of
$d=a\sqrt{2}$.  Hence, the solution in terms of ($\mathbf{E},\mathbf{B}$) in
this sytem of coordinates will look like (we assume $C>0$, and
$\varepsilon=-1$, $b(a,a)=3\frac{\pi}{4}$ and we write just $b$ for
$b(x,y)$).
\begin{multline}
\mathbf{E}=\biggl\{
C\frac{\theta(x,y;r_o)\theta(\xi-z;l_o)}
{{\rm cosh}\left[(x-a)^2+(y-a)^2\right]}
{\rm cos}\Bigl(\frac{z}{l_o}+b\Bigr);
\\[0.6ex]
C\frac{\theta(x,y;r_o)\theta(\xi-z;l_o)}
{{\rm cosh}\left[(x-a)^2+(y-a)^2\right]}
{\rm sin}\Bigl(\frac{z}{l_o}+b\Bigr);\ 0\biggr\}
	\end{multline}
	\begin{multline}
\mathbf{B}=\biggl\{
-C\frac{\theta(x,y;r_o)\theta(\xi-z;l_o)}
{{\rm cosh}\left[(x-a)^2+(y-a)^2\right]}
{\rm sin}\Bigl(\frac{z}{l_o}+b\Bigr);
\\[0.6ex]
C\frac{\theta(x,y;r_o)\theta(\xi-z;l_o)}
{{\rm cosh}\left[(x-a)^2+(y-a)^2\right]}
{\rm cos}\Bigl(\frac{z}{l_o}+b\Bigr);\ 0\biggr\}
	\end{multline}

We consider now the vectors $\mathbf{E}$ and $\mathbf{B}$ on the screw line
passing through the point $(a,a,0)$, where $b(a,a)=3\frac{\pi}{4}$. We choose
the coordinates $x,y,z$ as usual: $x$ grows  rightwards, $y$ grows  upwards,
and then $z$ is determined by the requirement to have a right orientation. If
we count the phase anticlockwise from the $x$-axis, then at $\xi=ct=0$, at
the point $P=(a,a,z=0)$ the vector $\mathbf{B}$ is directed to the point
$(0,0,0)$ and $\mathbf{E}$ is orthogonal to $\mathbf{B}$ in such a way that
$(\mathbf{E}\times\mathbf{B})$ is directed to $+z$. When $z$ grows from $0$
to $2\pi l_o$ the magnetic vector $\mathbf{B}$ on this screw line ( which is
the central screw line of the screw cylinder) turns around the axis $z$ and
stays always directed to it, while the electric vector $\mathbf{E}$  rotates
around the $z$-axis and it is always tangent to this rotation; at the same
time the Poynting vector keeps its  $+z$-orientation.  This situation
corresponds to the clockwise polarizarion (looking from behind).  If
$\kappa=-1$ then the electric vector $\mathbf{E}$ stays always directed to
the $z$-axis and since we have chosen $\varepsilon=-1$, i.e. the solution
propagates along $+z$, the magnetic vector $\mathbf{B}$ rotates around the
$z$-axis anticlockwise (looking from behind) and is always tangent to this
rotation.  So, this corresponds to the anticlockwise polarization.

We note once again that any actual assumption
$b(x,y)=const$ would make the representatives of the magnetic vector (when
$\kappa=1$) $\mathbf{B}$ parallel to each other at every point of the plane
$z=z_o,\ z_1<z_o<z_2$ at $t=0$. This may cause some instabilites with time,
so in order to have all representatives of $\mathbf{B}$ (or $\mathbf{E}$ in
the other polarization) at every point of that plane to be directed to the
$z-$axis and not to be paralel to each other, we may use $b(x,y)$ for the
necessary corrections.  This remark shows the importance of the relation
$b(x,y)\neq const$, $b(x,y)$ must be equal to the angle between the two lines
passing through the points $[(0,0,z_o);(a,a,z_o)]$ and
$[(0,0,z_o);(x,y,z_o)]$, where $x^2+y^2\leq r_o^2$.

Now, every point $(x,y,z)$ inside the solution region follows its own screw
line so, that the distance $\rho=\sqrt{x^2+y^2}$ between this screw line and
the $z$-axis is kept the same when $z$ grows. Since the spatial periodicity
along $z$ is equal to $2\pi l_o$, (note that $l_o$ is in fact the maximum
value of $l(x,y)$), we obtain an interpretation of the scale factor $L=l_o$
as the distance between the screw line inside $\Omega$ along which $L=l_o$
and the $z$-axis. This, in particular, means that for $r_o<<a\sqrt{2}$, i.e.
for a very thin screw cylinder, the scale factor $L=l(x,y)$ is approximated
by $L=l_o\theta(x,y;r_o)$; it also follows that the relation between the
longitudinal ant transverse sizes of the solution is approximately equal to
$\pi$.


\section{The Spin-momentum}
We turn now to the integral spin-momentum (helicity) computation. According
to our assumption its density is given by any of the correspondingly
normalized two 3-forms $F\wedge \delta F$ , or $(*F)\wedge (\delta *F)$.  In
order to have the appropriate physical dimension we consider now the 3-form
$\beta$ defined by
\begin{equation}
\beta=2\pi\frac{L^2}{c}F\wedge\delta F=
2\pi\frac{L^2}{c}\left[-\varepsilon
\Phi^2(\psi_\xi-\varepsilon \psi_z)dx\wedge dy\wedge
dz-\Phi^2(\psi_\xi-\varepsilon\psi_z)dx\wedge dy\wedge d\xi\right].
\end{equation}
Its physical dimension is "energy-density $\times $ time". Since $L=L(x,y)$
at most, and in view of (8), we see that $\beta$ is closed:
$\mathbf{d}\beta=0$.  The restriction of $\beta$ to ${\cal R}^3$ (which will
be denoted also by $\beta$) is also closed:
$$
\mathbf{d}\beta=-2\pi\mathbf{d}\left[\frac{L^2(x,y)}{c}
\varepsilon\Phi^2(\psi_\xi-\varepsilon \psi_z)dx\wedge dy\wedge dz\right]=0,
$$
and we may use the Stokes' theorem.  We shall make use of the solutions $3^o$
of equation (10). We have  $\psi_\xi =0,
\psi_z=\kappa /l(x,y)$, $L=|\psi_\xi-\varepsilon\psi_z|^{-1}=l$, so, in view
of our approximating assumption $L=l_o$, we integrate
$$
\beta=\frac{2\pi l_o}{c}\kappa\Phi^2 dx\wedge dy\wedge dz
$$
over the 3-space and obtain
\begin{equation}
\int_{{\cal R}^3}{\beta}=\kappa E\frac{2\pi l_o}{c}=
\kappa ET=\pm ET,
\end{equation}
where $E$ is the integral energy of the solution, $T=2\pi l_o/c$ is the
intrinsically defined time-period, and $\kappa=\pm 1$ accounts for the two
polarizations.  According to our interpretation this is the integral
intrinsic angular momentum, or spin-momentum, of the solution, for one period
$T$. This intrinsically defined action $ET$ of the solution is to be
identified with the Planck's constant $h$, $h=ET$, or $E=h\nu$, if we are
going to interpret the solution as an extended model of a single photon.
\vskip 0.3cm
\noindent
{\bf Remark}. For the connection of $F\wedge \delta F$ with our earlier
definition of the local intrinsic spin-momentum through the Nijenhuis
torsion tensor of $F^\nu_\mu$ one may look in our paper, cited as the last
one in [5].

\section{Conclusion}
According to our view, based on the conservation properties of the
energy-momentum and spin-momentum, photons are real objects and NOT
theoretical imagination.  This view on these microobjects as {\it real
extended} objects, obeying the famous Planck relation $E=h\nu$, almost
necessarily brings us to favor the soliton concept as the most appropriate
and self-consistent working theoretical tool for now, because {\it no
point-like conception is consistent with the availability of frequency and
spin-momentum of photons, as well as with the possibility photons to be
destroyed}.  So, the dynamical structure of photons is a real thing, clearly
manifesting itself through a consistent rotational-translational propagation
in space, and their finite nature reveals itself through the finite 3-volumes
of definite shape they occupy at every moment of their existence, and through
the finite values of the universal conserved quantities they carry.

The dynamical point of view on these objects reflects theoretically in the
possibility to make use of, more or less, {\it arbitrary} initial
configurations, i.e. to consider them as dynamical systems. This important
moment allows the localizing functions $\theta(x,...)$ from differential
topology to be used for making the spatial dimensions of the solution FINITE,
and NOT smoothly vanishing just at infinity as is the case of the usual
soliton solutions.

Our theoretical screw example, presenting an exact solution of the nonlinear
vacuum EED equations, was meant to present a, more or less, {\it visual
image} of the well known properties of photons, of their
translational-rotational dynamical structure, and, especially, of the nature
of their spin-momentum.  Of course, we do not insist on the function $1/{\rm
cosh}(.....)$ chosen, any other localisable inside the circle $A_{r_o}$
function would do the same job in this theoretical example. So far we have not
experimental data concerning the shape of single photons, and we do not
exclude various ones, so our choices for the amplitude $\Phi$ and for the
scale factor $L(x,y)$ are rather admissible approximations than correct
mathematical images.  The more important moment was to recognize the
dynamical sense of the quantity $F\wedge\delta F$, and to find that the
spin-momentum conservation equation $\mathbf{d}(F\wedge\delta F)=0$ gives
solutions for the phase function $\varphi={\rm cos}\psi$, which helps very
much to visualize the dinamical properties of photons in our aproach.  So, we
are able to obtain a general expression for the integral spin-momentum,
which, in fact, is the Planck's formula $h=ET$. Moreover, the whole solution
$(F,*F)$ describes naturally the polarization properties of photons, it
clearly differs the clockwise and anticlockwise polarizations through
pointing out the different roles of $\mathbf{E}$ and $\mathbf{B}$ in the two
cases.

An important moment in our approach is that $F$ and $*F$ are considered as
two components of the same solution, so the couples $(\mathbf{E},\mathbf{B})$
and $(-\mathbf{B},\mathbf{E})$ give together the 3-dimensional picture of
{\it one} solution. This, of course, would require the full energy-momentum
tensor ${\cal Q}_{\mu}^\nu$, to be two times the usual $Q_\mu^\nu$, given by
(4):  ${\cal Q}_\mu^\nu(F,*F)=2Q_\mu^\nu(F)=2Q_\mu^\nu(*F)$.

Finally, we'd like to mention that, making use of the localizing functions
$\theta(x,y,z;...)$, we may choose an amplitude function $\Phi$ of a
"many-lump" kind, i.e.  at every moment $\Phi$ to be different from zero
inside many non-overlaping 3-regions, probably of the same shape, so we are
able to describe a flow of consistently propagating photon-like exitations of
the same polarization and of the same phase. Some of such "many-lump"
solutions (i.e. flows of many 1-soliton solutions) may give the
macroimpression of, or to look like as, (parts of) plane waves.
\vskip 1.5cm
{\bf References}
\vskip 0.6cm
1. {\bf S.J.Farlow}, Partial Differential Equations for Scientists and
Engineers, John Wiley and Sons, Inc., 1982.

2. {\bf G.N.Lewis}, Nature {\bf 118} (1926), 874

3. {\bf J.J.Thomson}, Philos.Mag.Ser. 6, 48, 737 (1924), and 50, 1181
(1925), and Nature, vol.137, 23 (1936); {\bf N.Rashevsky}, Philos.Mag. Ser.7,
4, 459 (1927); {\bf W.Honig}, Found.Phys. 4, 367 (1974); {\bf G.Hunter,
R.Wadlinger}, Phys.Essays, vol.2,158 (1989), {\bf T.Waite}, Annales de la
Fondation Louis de Broglie, 20, No.4, (1995), {\bf J.P.Vigier}, Found. of
Physics, 21, 125, (1991), {\bf B.Lehnert}, Photon and Old Problems in Light
of New Ideas, Ed.V.Dvoeglazov, Nova Science Publ., 2000, pp.3-20.

4. {\bf S.Donev, M.Tashkova}, Proc.R.Soc.of Lond., A 443, 301, (1993);
 {\bf S.Donev, M.Tashkova},  Proc.R.Soc. of Lond., A 450, 281, (1995);
{\bf S.Donev, M.Tashkova}, Annales de la Fondation Louis de Broglie,
vol.23, No.2, 89-97, LANL e-print: nlin/9709006; vol.23, No.3-4, 103-115
(1998), LANL e-prints: nlin/9710001, 9710003.

5. {\bf R.Rajaraman}, Solitons and Instantons, North Holland, Amsterdam-New
York-Oxford, 1982

6. {\bf B.Dubrovin, S.Novikov, A.Fomenko}, Sovremennaya Geometriya (in
russian), sec. ed., Moskva, Nauka, 1986.

\end{document}